\documentclass[fleqn,usenatbib]{mnras}

\usepackage[T1]{fontenc}
\usepackage{ae,aecompl}

\usepackage{graphicx}	
\usepackage{amsmath}	
\usepackage{amssymb}	
\usepackage{makecell}
\usepackage{xcolor}

\newcommand{\msun}{M$_\odot$}
\newcommand{\msunyr}{\msun\,yr$^{-1}$}
\newcommand{\tento}[1]{10$^{#1}$}

\title[VLT/SPHERE sees dust around Antares]{The inner circumstellar dust of the red supergiant Antares as seen with VLT/SPHERE/ZIMPOL \thanks{Based on observations made with ESO Telescopes at the La Silla Paranal Observatory under programme ID 095.D-0458}}

\author[E. Cannon et al.]{
E. Cannon$^{1}$\thanks{E-mail: emily.cannon@kuleuven.be},
M. Montarg{\`e}s$^{1}$,
A. de Koter$^{1, 2}$,
L. Decin$^{1, 3}$,
M. Min$^{2, 4}$,
E. Lagadec$^{5}$, \newauthor
P. Kervella$^{6}$,
J.O. Sundqvist$^{1}$,
and H. Sana$^{1}$
\\
$^{1}$Institute of Astronomy, KU Leuven, Celestijnenlaan 200D B2401, 3001 Leuven, Belgium\\
$^{2}$Anton Pannekoek Institute of Astronomy, University of Amsterdam, The Netherlands\\
$^{3}$University of Leeds, School of Chemistry, Leeds LS2 9JT, United Kingdom\\
$^{4}$SRON Netherlands Institute for Space Research, Sorbonnelaan 2, 3584 CA Utrecht, The Netherlands\\
$^{5}$Laboratoire Lagrange, UNSA, CNRS, Obs.de la Cote d'Azur, Bd de l'Observatoire, 06304 Nice Cedex 4, France \\
$^{6}$LESIA, Observatoire de Paris, Universit{\`e} PSL, CNRS, Sorbonne Universit{\`e}, Universit{\`e} de Paris, 5 Place Jules Janssen,\\ 92195 Meudon, France }

\date{Accepted 2020 December 22. Received 2020 December 22; in original form 2020 July 30}

\pubyear{2020}

\begin{document}
\label{firstpage}
\pagerange{\pageref{firstpage}--\pageref{lastpage}}
\maketitle

\begin{abstract}
The processes by which red supergiants lose mass are not fully understood thus-far and their mass-loss rates lack theoretical constraints. The ambient surroundings of the nearby M0.5 Iab star Antares offers an ideal environment to obtain detailed empirical information on the outflow properties at its onset, and hence indirectly, on the mode(s) of mass loss. We present and analyse optical VLT/SPHERE/ZIMPOL polarimetric imaging with angular resolution down to 23 milli-arcsec, sufficient to spatially resolve both the stellar disk and its direct surroundings. We detect a conspicuous feature in polarised intensity that we identify as a clump containing dust, which we characterise through 3D radiative transfer modelling.  The clump is positioned behind the plane of the sky, therefore has been released from the backside of the star, and its inner edge is only 0.3 stellar radii above the surface. The current dust mass in the clump is $1.3^{+0.2}_{-1.0} \times 10^{-8}$ M$_{\odot}$, though its proximity to the star implies that dust nucleation is probably still ongoing. The ejection of clumps of gas and dust makes a non-negligible contribution to the total mass lost from the star which could possibly be linked to localised surface activity such as convective motions or non-radial pulsations.

\end{abstract}




\section{Introduction}

During their red supergiant (RSG) phase of evolution, massive stars (8 $<$ M $<$ 30 \msun; \citealt{2012A&A...537A.146E}) experience an important mass loss (\tento{-6} - \tento{-4}\msunyr; \citealt{2006ASPC..353..211V}; \citealt{2010A&A...523A..18D}) which strongly impacts their final mass, hence the properties of the supernova (SN) progenitor and of the compact remnant that is left behind. Material that is lost in the stellar wind, together with that ejected in the final core collapse, contributes to the chemical enrichment of the interstellar medium. The mass-loss properties of RSGs are however poorly constrained, and little is known about the mechanism(s) driving material from the surface. Without this knowledge it is difficult to build reliable theoretical models to predict mass-loss rates and therefore difficult to deduce the initial mass of SN IIP and IIL progenitors. This situation is problematic, not only from the perspective of understanding massive star evolution and the role of RSG in stellar feedback, but also with respect to fathoming the so-called Red Supergiant Problem (i.e. the lack of detections of progenitors with initial mass $>$ 17\,\msun; \citealt{2009ARA&A..47...63S}), which may be connected to an underestimated RSG mass-loss (\citealt{2012A&A...537A.146E}; \citealt{2012A&A...538L...8G}).

Several mechanisms have been proposed that could contribute to RSG mass loss. Most invoke a stellar wind driven by radiation pressure on dust grains, similar to the mechanism proposed for Asymptotic Giant Branch (AGB) stars (\citealt{2008A&A...491L...1H}; \citealt{2016A&A...594A.108H}). However, it is doubtful whether the prevailing thermodynamic conditions to grow grains in the first place are comparable for AGB and RSG stars \citep{2019A&A...623A.158H}. Radiation pressure exerted on spectral lines of molecular species may be an alternative, though at the present time it is unclear whether this provides sufficient driving in the onset region of the flow \citep{2010ASPC..425..181B}. The possibility of an Alfv\'{e}n wave-driven wind has also been explored by \citet{2000ApJ...528..965A} and a wave-driven mass-loss model for Betelgeuse can be seen in \citet{1984ApJ...284..238H}. A further alternative is that the mass-loss trigger is linked to surface activity, where pulsations and large convective cells upwelling from the sub-photosphere may lower the effective gravity allowing radiation pressure to launch material \citep{2007A&A...469..671J}.


Observations of Antares, one of the closest RSGs, by \citet{2017A&A...605A.108M} with the VLTI/PIONIER instrument revealed that convective cells of various sizes cover the stellar surface, confirming early indications of the presence of such structures by  \citet{1997MNRAS.285..529T} who detected variable hot spots on the stellar surface and \citet{1990A&A...230..355R} who suspected an asymmetric brightness profile. Above the surface (out to 1.7 stellar radii) turbulent motion of large clumps of gas were observed using VLTI/AMBER \citep{2017Natur.548..310O}. Moving further away from the star, \citet{2014A&A...568A..17O} detected large clumps containing dust within 40\,$-$\,96 R$_{\star}$ with VLT/VISIR, upholding first indications for the presence of dusty clumps by \citet{2001ApJ...548..861M}. To investigate whether there is a link between surface convection and these clumps in the ambient environment, observations of the innermost circumstellar environment are needed. This is made possible by the high spatial resolution polarimetry capabilities of SPHERE/ZIMPOL at ESO's VLT observatory in Paranal, Chile.
This instrument, with angular resolution up to 23 mas is capable of resolving the surfaces of the two closest RSGs, Antares and Betelgeuse, allowing the dust in the inner wind to be probed in detail. Betelgeuse has previously been observed using this instrument by \citet{2016A&A...585A..28K}. 

In this paper we present SPHERE/ZIMPOL observations of the RSG Antares along with 3D radiative transfer modelling of its ambient surroundings in order to characterise the spatial distribution and amount of dust near the surface.
The observations and data reduction are described in Sect. \ref{sec:obs} followed by their analysis in Sect. \ref{sec:analysis}. The model setup and results are described in Sect. \ref{sec:modelling}. We discuss our findings in Sect. \ref{sec:discussion} and end with a summary and conclusions in Sect.~\ref{sec:summary}.


\section{Observations and data reduction}
\label{sec:obs}

Antares ($\alpha$\,Sco A, HD\,148478, HD\,6134) is an M0.5\,Iab  \citep{1984ApJS...55..657C} RSG star at a distance of $170^{+29}_{-25}$\,pc \citep{2007A&A...474..653V}. As such, it is one of the largest and visually brightest stars in the sky. It has an angular diameter of $37.89 \pm 0.10$ mas at the H$^{-}$ opacity minimum at 1.61\,$\mu$m \citep{2017A&A...605A.108M} and $37.38 \pm 0.06$\,mas in the K-band continuum \citep{2013A&A...555A..24O}. Antares is known to have a companion, the B2.5\,V star $\alpha$\,Sco\,B, at a 2.73" angular separation (in 2006), approximately 224\,AU behind the supergiant \citep{2008A&A...491..229R}.

Observations of Antares, and a corresponding point spread function (PSF) calibrator star, $\epsilon$ Sco, were taken on 25 June 2015. These observations were carried out using SPHERE/ZIMPOL  \citep{2019A&A...631A.155B}, a high resolution adaptive optics imaging polarimeter, at ESO's Very Large Telescope (VLT). Antares and its calibrator were observed in six filters in the visible. The log of the observations is presented in Table \ref{obstab}. The filter characteristics are given in Table \ref{fluxtab}.
\begin{table*}
\caption{Log of SPHERE/ZIMPOL observations of Antares and reference star $\epsilon$ Sco.}             
\label{table:1}      
\centering          
\begin{tabular}{l l l c c c c c }     
\hline       
                     
Star & \thead{Time UT \\2015-06-25} & Filter & ND & $\theta$ ["] & DIT[s] $\times$ NDIT & AM  & $\theta_\text{PSF}$ [mas] \\ 
\hline  \noalign{\vskip 2mm}

      Antares  & 01:29:08  & CntH$\alpha$  &  ND\_2   &  0.61   &  1.2 $\times$ 20     &  1.069    & - \\
        &  & NH$\alpha$    &  ND\_2   &  0.61   &  1.2 $\times$ 20     &  1.069    & - \\
        & 01:55:12  & BH$\alpha$    &  ND\_2   &  0.61   &  1.2 $\times$ 20     &  1.034    & - \\
        &  & TiO           &  ND\_2   &  0.61   &  1.2 $\times$ 20     &  1.034    & - \\
        & 02:27:17 & V           &  ND\_1   &  0.66   &  1.2 $\times$ 20     &  1.009    & - \\
        &   & KI            &  ND\_1   &  0.66   &  1.2 $\times$ 20     &  1.009    & - \\
        \noalign{\vskip 2mm}

       $\epsilon$ Sco & 00:40:42 & CntH$\alpha$  &  ND\_1    &  0.70   &  1.2 $\times$ 20  & 1.236 &  26 \\
        &  & NH$\alpha$    &  ND\_1    &  0.70   &  1.2 $\times$ 20  & 1.236 &  27 \\
        & 00:53:29 & BH$\alpha$    &  ND\_1    &  0.83   &  1.2 $\times$ 20  & 1.197 &  25 \\
        &  & TiO           &  ND\_1    &  0.83   &  1.2 $\times$ 20  & 1.197 &  24 \\
           & 01:05:17 & V           &  ND\_1    &  0.61   &  1.2 $\times$ 20  & 1.166 &  30 \\
        &  & KI            &  ND\_1    &  0.61   &  1.2 $\times$ 20  & 1.166 &  25 \\
    \hline\noalign{\vskip 2mm}

\end{tabular}
\label{obstab}
\vspace{1ex}\\
\begin{flushleft}
\textbf{Note}: ND indicates which neutral density filter has been used, $\theta$ is the visible seeing, AM gives the airmass and $\theta_\text{PSF}$ is the FWHM of the PSF images. DIT gives the integration time of each frame and NDIT is the number of integrations. The filter pairs grouped together in time were observed simultaneously using the two arms of the detector. The characteristics of the filters are given in Table \ref{fluxtab}.
\end{flushleft}
\end{table*}

\begin{table}
\caption{Filter characteristics and calculated photometry of Antares.}           
\centering                       
\begin{tabular}{l l l c}      
\hline              
Filter & $\lambda$ [nm] & $\Delta\lambda$ [nm] & \thead{Flux \\ $10^{-8}$ W m$^{-2}$ $\mu$m$^{-1}$} \\  
\hline  \noalign{\vskip 2mm}                     
   V                & 554       & 80.6  & 1.555 $^{+0.642}_{-0.561}$\\ 
      \noalign{\vskip 1.2mm}     
   CntH$\alpha$     & 644.9     & 4.1   & 2.894 $^{+1.003}_{-0.844}$ \\
      \noalign{\vskip 1.2mm}    
   BH$\alpha$       & 655.6     & 5.5   & 2.556 $^{+0.852}_{-0.722}$\\ 
      \noalign{\vskip 1.2mm}    
   NH$\alpha$       & 656.34    & 0.97  & 2.964 $^{+0.968}_{-0.831}$ \\
      \noalign{\vskip 1.2mm}    
   TiO              & 716.8     & 19.7  & 2.545 $^{+0.794}_{-0.689}$\\ 
      \noalign{\vskip 1.2mm}    
   KI               & 770.2     & 21.2  & 3.679 $^{+1.053}_{-0.922}$\\
      \noalign{\vskip 1.2mm}    
\hline                                   
\end{tabular}
\label{fluxtab}
\vspace{1ex}\\
\begin{flushleft}
\textbf{Note}: $\Delta\lambda$ indicates the FWHM of the filter. 
\end{flushleft}
\end{table}

\begin{figure*}
      \includegraphics[width = 0.87\textwidth]{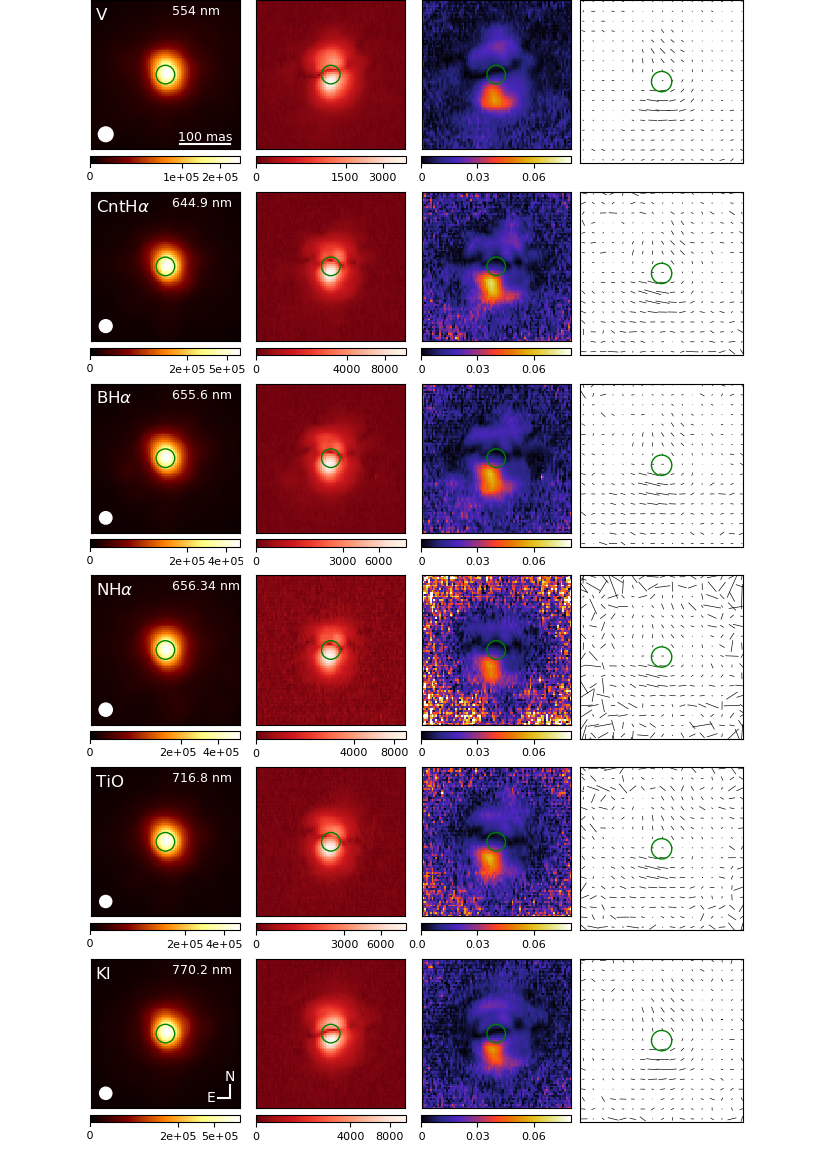}
      \caption{First column: Intensity images (W m$^{-2}$ $\mu$m$^{-1}$ sr$^{-1}$) of Antares, in square root scale. The filled white circle indicates the beam size and the green circle indicates the size of the photosphere \citep{2013A&A...555A..24O}. Second column: Polarised flux (W m$^{-2}$ $\mu$m$^{-1}$ sr$^{-1}$), in square root scale. Third column: Degree of linear polarisation in linear scale spanning 0 - 8\%. Last column: Angle of the polarisation vector. The magnitude of the vector is scaled to the strength of the degree of linear polarisation at each point. 
              }
         \label{observations}
   \end{figure*}

\begin{figure*}
      \includegraphics[width = 0.87\textwidth]{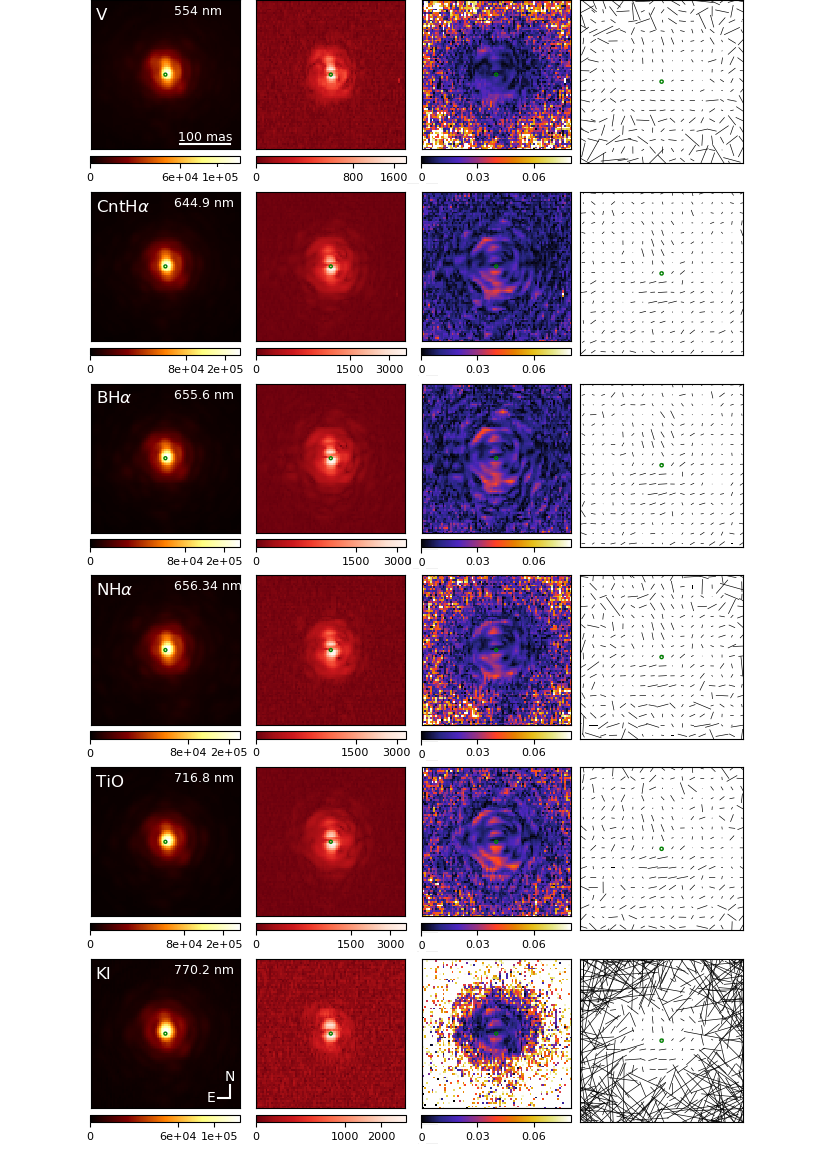}
      \caption{Same as Fig. \ref{observations} but for $\epsilon$ Sco.}
         \label{psf}
   \end{figure*}

Due to atmospheric or instrumental conditions, the adaptive optics loop of the instrument frequently opened during the observations. To eliminate the corrupted frames, we performed a selection on the raw data before executing the instrument pipeline. We determined the average of the flux of each frame in a circular area centred on the maximum. For each couple target/filter, we rejected the frames for which the mean central flux was below a specific threshold. The threshold in each case is then defined by 0.5, 2 or 5 standard deviations from the mean depending on the dispersion of the flux points for the individual case. Next, the raw data were processed using the ESO reflex data reduction pipeline v0.24.0 \citep{2013A&A...559A..96F}. From this data we then computed the total intensity, the degree of linear polarisation (DoLP), the polarised intensity and the polarisation position angle as described by \citet{2015A&A...578A..77K}. The linear polarised intensity (Fig. \ref{observations}) is defined by $\sqrt{Q^2 + U^2}$ where Q and U are Stokes parameters. The DoLP, being the polarised intensity divided by the intensity, gives us the fraction of the light that is linearly polarised. The observations were centred by fitting a Gaussian function to the intensity images to locate the centre of the star. The total intensity images of Antares were deconvolved using the Lucy - Richardson deconvolution algorithm implemented in IRAF \citep{1974AJ.....79..745L} using the intensity images of $\epsilon$ Sco as the PSF (Fig. \ref{psf}). The number of iterations needed in this process was judged by matching the measured angular diameter of the star to that of the deconvolved image. Beyond five iterations no further changes were measured in the full width half maximum (FWHM) of the intensity profiles.

\begin{figure*}
      \includegraphics[width = \textwidth]{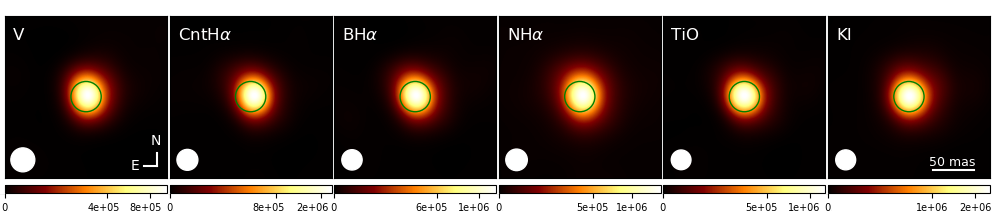}
      \caption{The deconvolved intensity images (W m$^{-2}$ $\mu$m$^{-1}$ sr$^{-1}$) of Antares in each filter, shown in square root scale. The photospheric size, measured by \citet{2013A&A...555A..24O} in the near-infrared, is shown in green.}
         \label{decon}
   \end{figure*}

The SPHERE/ZIMPOL pipeline products are not flux calibrated. In order to derive the observed flux in each filter for Antares we first matched a stellar atmosphere model from \citet{2003IAUS..210P.A20C} to the calibrator star, $\epsilon$ Sco. As $\epsilon$ Sco is a K1III star \citep{2006AJ....132..161G}, we selected a model with an effective temperature of 4500K, log g = +2.50 with solar metallicity in agreement with \citet{1990ApJS...74.1075M}. We scaled this theoretical spectral energy distribution (SED) using the angular diameter of the star, 5.747 $\pm$ 0.008 mas \citep{2009MNRAS.399..399R}, and  compared to existing photometry \citep{2002yCat.2237....0D} to confirm the suitability of the model. The theoretical flux for each of the SPHERE filters which were observed was calculated by integrating over the SED after convolving with the SPHERE transmission filters. We determined the observed flux in the image by summing the intensity over a circular aperture of 184 mas centred on the star. The background flux was estimated using a ring, with inner and outer radii of 184 and 220 mas respectively, and subtracted. From this we obtained a conversion factor that was applied to the Antares data. The uncertainty estimation on the calibrated flux is dominated by that on the effective temperature of the calibrator star. This uncertainty is therefore set by the spacing of the stellar atmosphere model grid as this sampling is certainly larger than the uncertainty of the effective temperature leading to a conservative uncertainty estimate of $\pm$ 250K. The flux calibrated intensity images of Antares are represented in Fig. \ref{observations} together with the polarimetric maps. The same for $\epsilon$ Sco is shown in Fig. \ref{psf}. The deconvolved intensity images of Antares can be seen in Fig.\ref{decon}.

\section{Data analysis}
\label{sec:analysis}

\subsection{Intensity}

\begin{figure}   
   \centering
   \includegraphics[width=\hsize]{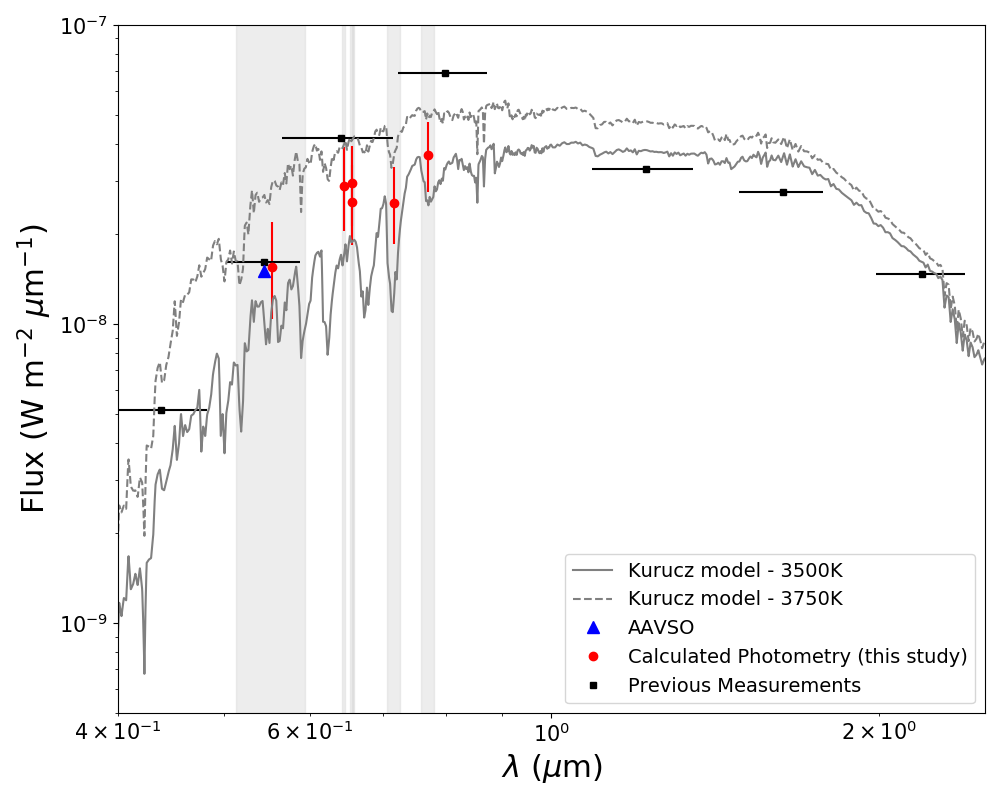}
      \caption{Antares photometry. A stellar atmosphere model from \citet{2003IAUS..210P.A20C} with T$_\text{eff} = 3500$\,K, log g = 0 and solar metallicity, scaled to the angular diameter of Antares and reddened is shown in grey. Similarly a model with T$_\text{eff} = 3750$\,K is shown by the dashed line. The grey shaded areas represent our six ZIMPOL filters (NH$\alpha$ and BH$\alpha$ overlap). The horizontal bars on the previously measured photometry by \citet{2002yCat.2237....0D} give the width of the Johnson bands. The blue triangle is a measurement in the V band from AAVSO that was taken on 26-06-2015 (one day apart from our observations).}
    \label{flux_cal}
   \end{figure}

Figure \ref{flux_cal} shows the flux obtained in each ZIMPOL filter we observed.  As Antares is a semi-regular variable we also compared our flux with a measurement from the American Association of Variable Star Observers (AAVSO) in the V band (Fig. \ref{flux_cal}) taken within 24 hours of our SPHERE observations and found our results to be in agreement. They are broadly consistent with previous measured photometry by \citet{2002yCat.2237....0D}.

With a beam size (see Table \ref{obstab}) for each of the observations smaller than that of the projected surface size of the star the stellar disk is resolved in all filters. Fitting a two dimensional Gaussian function to the deconvolved intensity images (Fig. \ref{decon}), reveals that the visible photosphere departs from spherical symmetry in all filters giving eccentricities between 0.4 and 0.52 which could perhaps indicate that there is a temperature variation on the surface of the star.

\subsection{Polarised flux}

Significant signal can be seen in the polarised flux and DoLP across the six filters (see 3$^{\rm rd}$ column in Fig.~\ref{observations}). The DoLP that we see in each of the images is considerably higher than the polarisation that is caused by the instrument itself, which is approximately 0.5\% \citep{2019A&A...631A.155B}.

There is a dark lane that runs through the centre  of all polarisation images which is nonphysical and is due to a beam shift effect. This effect is introduced by the instruments mirrors and is further detailed by \citet{2018A&A...619A...9S}. Outside of the area plotted in Fig. \ref{observations} and Fig. \ref{psf} the images are dominated by noise (causing the large polarisation vectors at the edges of the images; for the corresponding RMS maps see Fig. \ref{rms}). No significant stellar signal is seen in the DoLP of the calibrator star (Fig. \ref{psf}), the signal present in these images follows the diffraction rings caused by the telescopes mirror and is split East-West by the beam shift effect.

A large, conspicuous feature can be seen to the south of Antares in the DoLP in all filters. In the plane of the sky, the onset of the feature appears to be right at the stellar surface with a projected surface size greater than that of the star. This polarisation could be caused by the scattering of the light by circumstellar dust grains or scattering off molecules or free electrons  \citep{2000ApJS..128..245Z}. The latter option seems unlikely given the low ionisation temperature associated to the radiation field from Antares. The polarisation seen in the observations is consistent with the dust hypothesis as it is present throughout the entire wavelength range which would not be expected if the polarisation was a result of light scattering off molecules within specific lines. The directions of the polarisation vectors at the location of the feature are tangential to the stellar photosphere, as expected from the scattering of light from the central source by dust in the circumstellar environment. Henceforth, we refer to the feature as the clump.


\section{\textsc{MCMax} modelling}
\label{sec:modelling}

In order to determine if the dust hypothesis to explain the polarisation signal around Antares is plausible and to characterise the dust causing the scattering, we run a radiative transfer model using \textsc{MCMax3D} \citep{2009A&A...497..155M}, a 3-D Monte Carlo radiative transfer code. \textsc{MCMax3D} implements Mie scattering on a distribution of dust grains modelled as hollow spheres. The latter does not imply that the grains are truly hollow spheres; rather this assumption assures that the symmetrical nature of solid spherical particles, as in standard Mie theory, is broken \citep{2005A&A...432..909M,2004ldce.conf...28M}. An ensemble of hollow spheres thus better represents the properties of a true distribution of non-spherically symmetric particles. The images of the Stokes vectors (I, Q and U) are computed taking into account the diameter of the telescope mirror and can therefore be compared directly to the SPHERE observations. Through this modelling of our observations we aim to provide constraints on the spatial distribution and total mass of the dust. 

We focus the modelling on the conspicuous clump to the south of Antares that is seen in the DoLP. The polarisation signal to the north of the star is several times fainter than that in the south and not well defined in all filters, for these reasons we concentrate our modelling efforts on the southern clump. We approximate the clump as a sphere of dust with a constant density as the number of resolution elements over the clump is a few. Therefore we would have no diagnostic tools to constrain properties of a (spherical) density structure in the clump. The centre of the clump is placed at different ($x,y,z$) positions relative to the stellar centre, such that if we can constrain its 3D position we may link it to a surface release location assuming the clump is moving radially outwards. For clarity, $z$ is the position along the line of sight (positive in the direction of the observer), $y$ is the north-south axis (where north is positive), and $x$ the east-west axis (where west is positive). The $x$ coordinate of the clump centre is kept fixed based on the observations. Below we discuss the assumptions regarding the composition and size of the dust particles. If we ignore the latter properties for now, our modelling space consists of four variables: $z$ and $y$ position, radius $R_{\rm clump}$ and dust mass of the clump $M_{\rm clump}$. To probe these dimensions we have constructed a grid of models of varying step-size in each dimension. The ranges and step sizes of each of the four parameters are visualised in Fig.~\ref{chi2}. We note here, first, that $R_{\rm clump}$ was varied between 1$-$5 $R_{\star}$, where $R_{\star} = 680\,$R$_{\odot}$ is the radius of Antares. Second, the parameter controlling the mass of the dust is sampled rather unevenly to ensure an unbiased density sampling as the volume of the dust sphere varied. Third and finally, the $y$ coordinate was allowed to vary as it was difficult to determine how far south the centre of the clump is in the observations as the 
models show that the DoLP signal is not constant through the large clump, i.e. the
signal is concentrated in a smaller portion of the dust sphere (see also Sect.~\ref{sec:discussion}).

\subsection{Parameters and Assumptions}
\subsubsection{Stellar parameters}

To determine the stellar energy distribution to use as input for the radiative transfer models we first applied a reddening law to two stellar atmosphere models from \citet{2003IAUS..210P.A20C} with T$_\text{eff} = 3500$\,K and 3750\,K, and surface gravity $g = 1$\,cm\,s$^{-1}$. We use A$_V$ = 0.43 and R$_V$ = 3.1 from \citet{2013A&A...555A..24O} and follow the law described in \citet{1989ApJ...345..245C}. Both models are plotted in Fig. \ref{flux_cal} and show that the photometry we retrieved from the SPHERE data falls between the two models. As \citet{2013A&A...555A..24O} determines a T$_\text{eff}$ of 3660 $\pm$ 120\,K this is unsurprising. A $\chi^2$ test determines the model with T$_\text{eff} = 3500$\,K to be the better fit so we proceed with a stellar atmosphere model at this temperature for the \textsc{MCMax3D} modelling input. A test between two radiative transfer models with T$_\text{eff} = 3500$\,K and 3750\,K, showed no significant difference in the DoLP. The luminosity of 62,500 L$_\odot$ was set so that the angular diameter of the model star was consistent with previous interferometric measurements from \citet{2013A&A...555A..24O} at a distance of 170 pc.

The star is modelled as a spherical object though both simulations and observations of RSGs indicate that this is not exactly the case due to their large convective cells. Similarly, the code assumes isotropic light emission. For the scope of this study these assumptions are reasonable as we focus on the modelling of a large dust clump illuminated by a large fraction of the surface of the star.

\subsubsection{Dust composition and grain size}
\label{sec:dust_composition}

\citet{2017A&A...603A.116A} report that the chemical composition of circumstellar dust grains cannot be reliably determined using observations in the visible regime alone but also need measurements in the near-IR. We therefore do not attempt to do so here. \citet{2009A&A...498..127V} review studies addressing this topic and provide an overview of likely dust constituents: aluminium oxide (Al$_{2}$O$_{3}$); melilite (Ca$_{2}$Al$_{2}$SiO$_{7}$); olivine (Mg$_{2x}$Fe$_{2-2x}$SiO$_{4}$; $0 \leq x \leq 1$); iron magnesium oxide (MgFeO); metallic iron (Fe), and carbon (C). Aluminium oxide is expected to condense early on in the condensation cycle \citep[e.g.][]{1990fmpn.coll..186T}.
A study of dust precursors in asymptotic giant branch (AGB) star winds by \citet{2019MNRAS.489.4890B} finds Al$_2$O$_3$ to be a potential first species to condense from the gas phase on the condition that the monomer (Al$_2$O$_3$)$_{n=1}$ forms. Depending on local density, cluster formation starts at temperatures as high as 1600 $-$ 2200\,K the highest of all species trialled by them. 
As the dust in the observations appears close to the stellar surface and therefore at high temperatures, for this study we used dust composed of Al$_2$O$_3$ adopting the optical properties from \citet{1997ApJ...476..199B} derived in the range 7.8\,$-$\,200 $\mu$m. For shorter wavelengths, the optical constants are extrapolated following \citet{1998asls.book.....B}. This extrapolation implies that the alumina grains are almost transparent at optical wavelength, in line with findings by \citet{1995Icar..114..203K}. A study by \citet{2016A&A...594A.108H} explores Al$_2$O$_3$ formation around M-type AGB stars. In this study they show that Al$_2$O$_3$ forms closer to the star than silicates and may act as seed particles for the condensation of silicates further out.

Two other compositions were trialled, a mixture of MgSiO$_3$ + amorphous carbon, and the dust composition found by \citet{2009A&A...498..127V} comprising majorly of melilite with smaller amounts of olivine, alumina and carbon. We found that the DoLP is not strongly sensitive to composition, though the best fit parameters ($x$, $y$, $R_{\rm dust}$, $M_{\rm dust}$) will differ somewhat from those derived using aluminium oxide. Specifically, these other compositions produce slightly less polarisation. These differences are, however, so small that they do not impact our conclusions. We do point out that these alternative grains are more opaque than Al$_{2}$O$_{3}$, therefore their temperatures are higher. We established that if the grain composition is actually a mixture of species -- which is very likely -- part of the volume of the best solution for our aluminium oxide only model would be too hot for these other grains to exist, either as thermally isolated species or as species in thermal contact. For the \citet{2009A&A...498..127V} mixture this volume fraction is about 8 percent (10 percent when T$_\text{eff}$ = 3750\,K) if a condensation temperature of 1500\,K is assumed, a result that has a modest impact on the assumption for the dust-to-gas ratio of the clump (see Sect.~\ref{sec:cloud_properties}). One should note that this estimation does not take into account the presence of a warm chromosphere (see e.g. \citet{2001ApJ...551.1073H} and references therein) around a RSG such as Antares. Recently, \citet{2020A&A...638A..65O} have shown that the interaction between the warm chromosphere at several thousand Kelvin and the cool gas which allows dust condensation in the same location is very complex: the detection of one or another is highly dependent on the wavelength used for the observations. It is likely that the warm gas is not dense and has a limited role on the dust condensation sequence. Therefore, as \textsc{MCMax3D} cannot take this temperature profile into account, we do not include it in our models. In conclusion: we expect the composition (in the bulk of the clump) to be a mixture typical for RSG outflows, however, we adopt the optical properties of aluminium oxide in our modelling to avoid partial dust condensation issues in the clump. Derived dust masses represent that of the actual dust mixture.

The DoLP is sensitive to the size distribution of the grain population; large grains producing less polarisation relative to small grains in the wavelength range of our observations. Our observational data does not allow to place firm constraints on these properties, and we limit our investigation to assessing whether the wavelength dependence of the DoLP in the best fit model is consistent with our measurements (see Sect.~\ref{sec:modelling} and Fig.~\ref{dolp_wave}). We follow \citet{2009A&A...498..127V} and adopt an MRN distribution of sizes described in \citet{1977ApJ...217..425M}, $n(a) \propto a^{-3.5}$, characteristic for interstellar particles, with sizes $a$ in the range 0.01$-$1 $\mu$m. We prefer this approach over adopting a single particle size (e.g. \citealt{2001A&A...368..950V}, \citealt{2016MNRAS.463.1269B}) as likely stochastic processes play a role in dust formation and growth. Our adopted size distribution is shifted to slightly larger grains in comparison with \citet{2012ApJ...759...20K} and \citet{2014A&A...568A..17O}. Micron sized grain in the circumstellar environment of Antares are reported in \citet{1987ApJ...321..921S} giving weight to theoretical considerations of dust driving in cool star outflows which seem to favour fairly large grains \citep{2008A&A...491L...1H}.

\subsection{Modelling results}
\label{sec:modellingres}

To determine the best-fitting model to the observations we used a $\chi^2$ minimisation technique. First we compute $\chi^2$ for each of our models using:
\begin{equation}
\chi^2=\sum^{n}_{i=1} \frac{(O_i - E_i)^2}{\sigma_i^2}
\end{equation}
\noindent 
where O and E is the flux in pixel $i$ for our observations and models respectively, $\sigma$ is the error in the observed data (see Fig. \ref{rms}) and $n$ is the total number of pixels in our cubes ($n = 33620$). In this case we directly compared the DoLP image outputted by our models to the observed images in all filters by summing over the total number of pixels after cutting the images to a 300 $\times$ 300 mas field of view.

In order to calculate the confidence interval on the parameter ranges we use the same method as outlined in \citet{2011ApJ...741L...8T} and \citet{2019ApJ...880..115A}. The $\chi^2$ values from the grid were normalised such that the best-fitting model had a $\chi^2_{\text {red}}$ value of 1. From here the P-value was calculated, all models with a P-value higher than 0.05 (therefore within the 95\% confidence interval) are deemed acceptable models that are statistically indistinguishable from each other. It is from these models that we are taking our parameter ranges for the position, radius and dust mass of the clump. Due to computational limitations we had to limit the sampling rate. To account for this we perform a linear interpolation to the upper P-value points to give a better estimate on the confidence interval on each parameter.

\begin{figure*}
      \includegraphics[width = \textwidth]{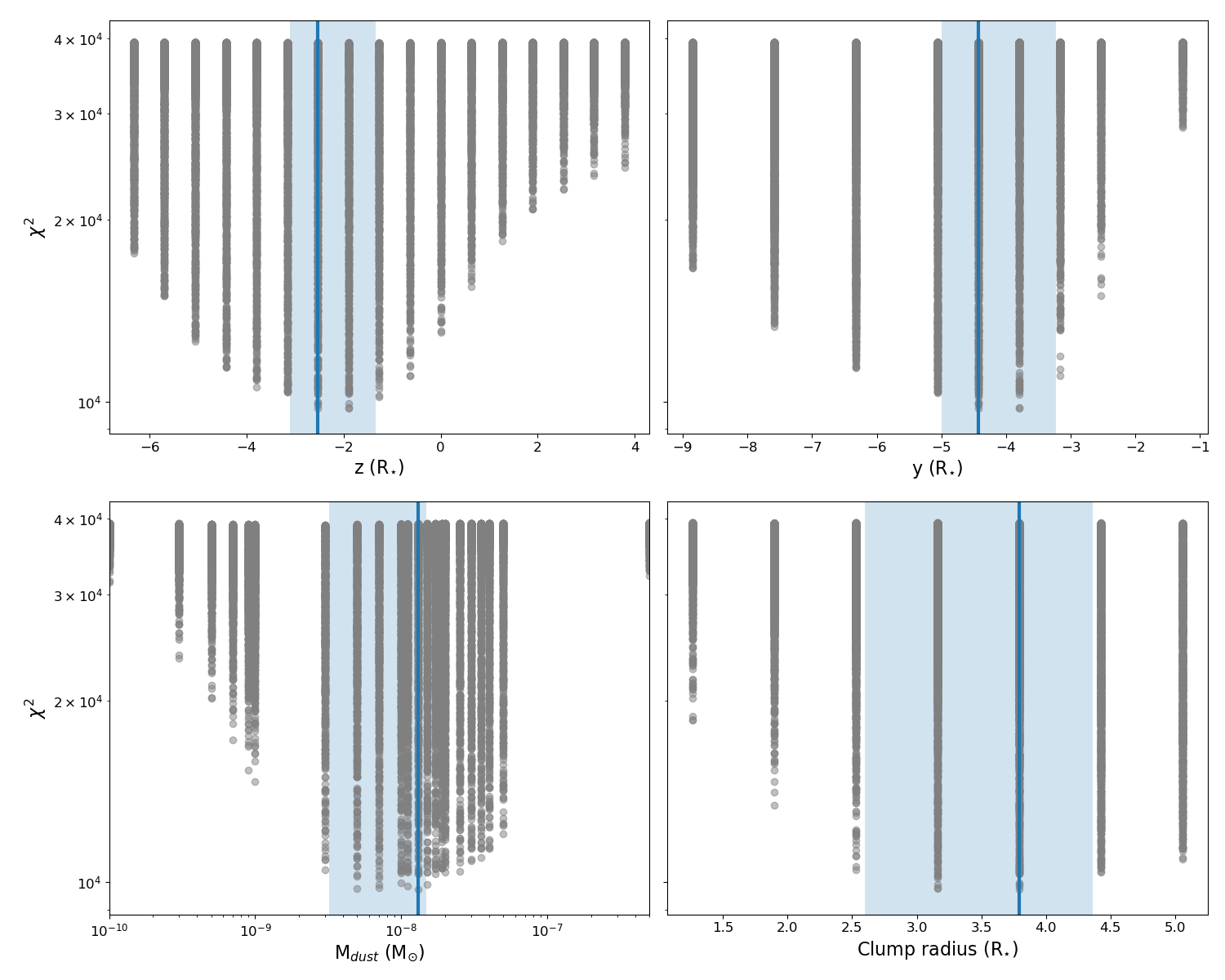}
      \caption{$\chi^2$ values derived from the comparison of the degree of linear polarisation from the ZIMPOL observations and MCMax3D simulations against the four variables. Here, $z$ is the line of sight and $y$ is the North-South axis. The solid line indicates where the best-fitting model falls and the shaded region shows the confidence intervals.}
         \label{chi2}
   \end{figure*}

Figure \ref{chi2} shows the $\chi^2$ distribution for our grid of models. For each of our variable parameters relatively clear minima can be seen in the $\chi^2$ distribution. Table \ref{model_params} shows the parameters of the best-fitting models from the grid using Al$_2$O$_3$ for the dust composition. As can be seen in Fig. \ref{chi2} all of these models place the dust behind the plane of the star at close proximity to the photosphere. Table \ref{model_params} shows that as the mass and radius of the dust clump display large confidence intervals, the dust density in the clump remains relatively constant for our best fitting models.

\begin{figure*}
      \includegraphics[width = 0.7\textwidth]{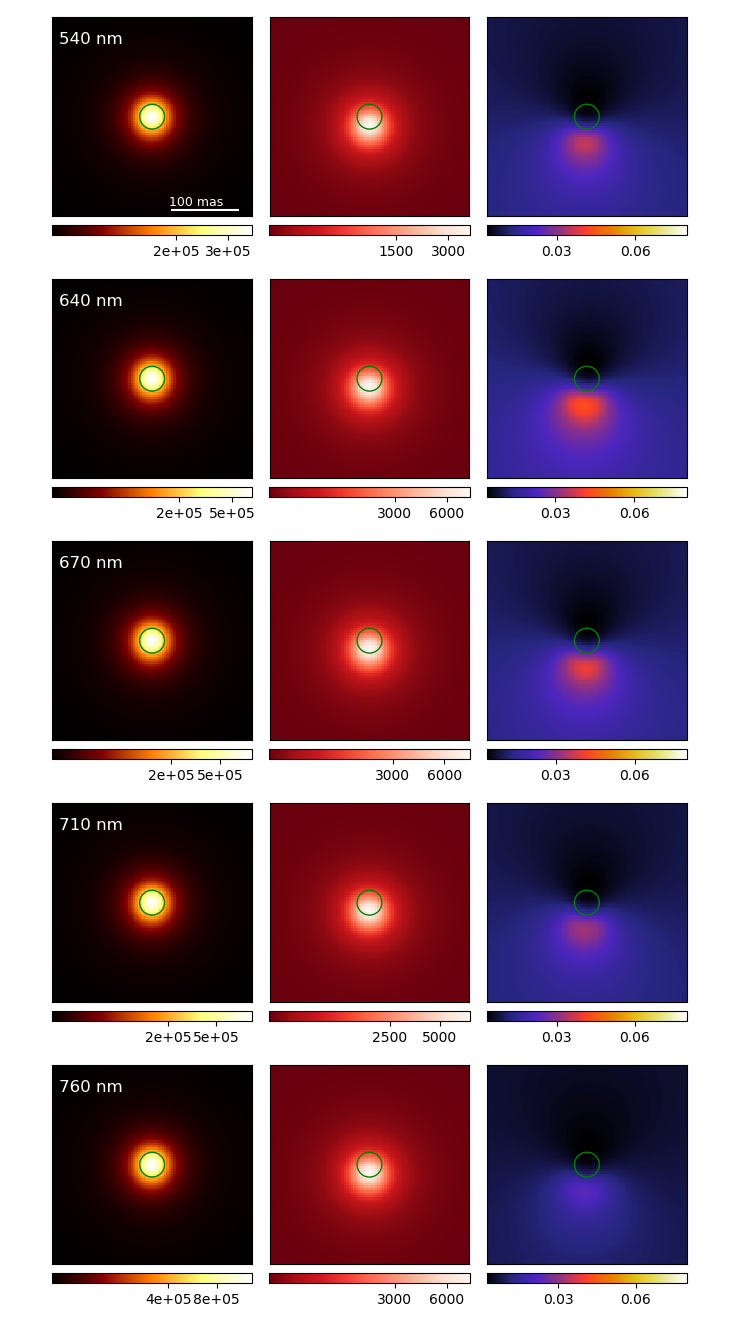}
      \caption{Best matched \textsc{MCMax3D} model as determined by comparison of the DoLP to the observations. First column: Intensity images (W m$^{-2}$ $\mu$m$^{-1}$ sr$^{-1}$). The green circle indicates the size of the photosphere. Second column: Polarised flux (W m$^{-2}$ $\mu$m$^{-1}$ sr$^{-1}$). Third column: Degree of linear polarisation. }
         \label{model}
   \end{figure*}

\begin{figure}   
   \centering
   \includegraphics[width=\hsize]{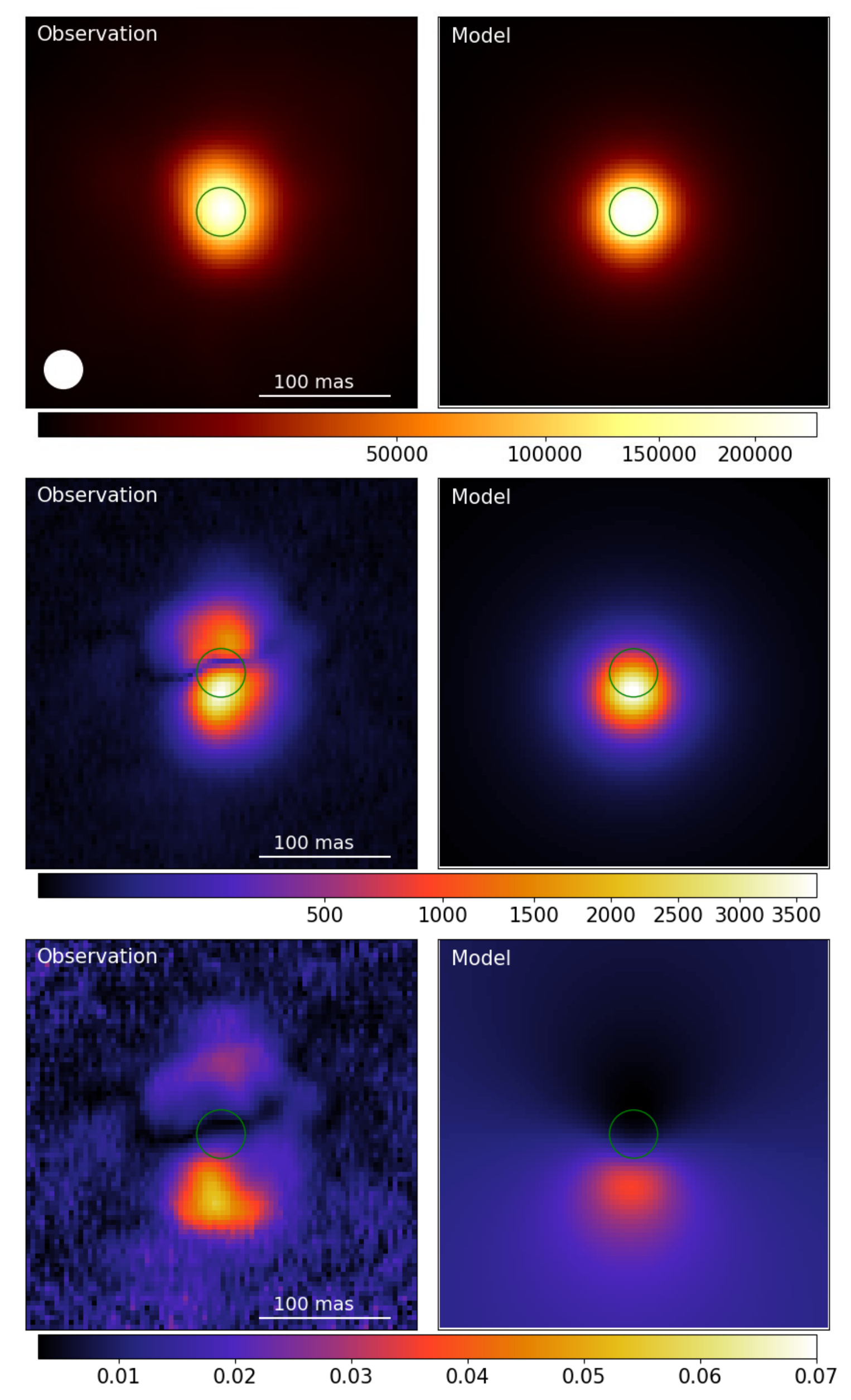}
      \caption{Comparison of observations (left) to the models (right) in the V filter. First row: Intensity images (W m$^{-2}$ $\mu$m$^{-1}$ sr$^{-1}$). The green circle indicates the size of the photosphere. Second row: Polarised flux (W m$^{-2}$ $\mu$m$^{-1}$ sr$^{-1}$). Third row: Degree of linear polarisation.}
    \label{sidebyside}
   \end{figure}

\begin{figure}   
   \centering
   \includegraphics[width=\hsize]{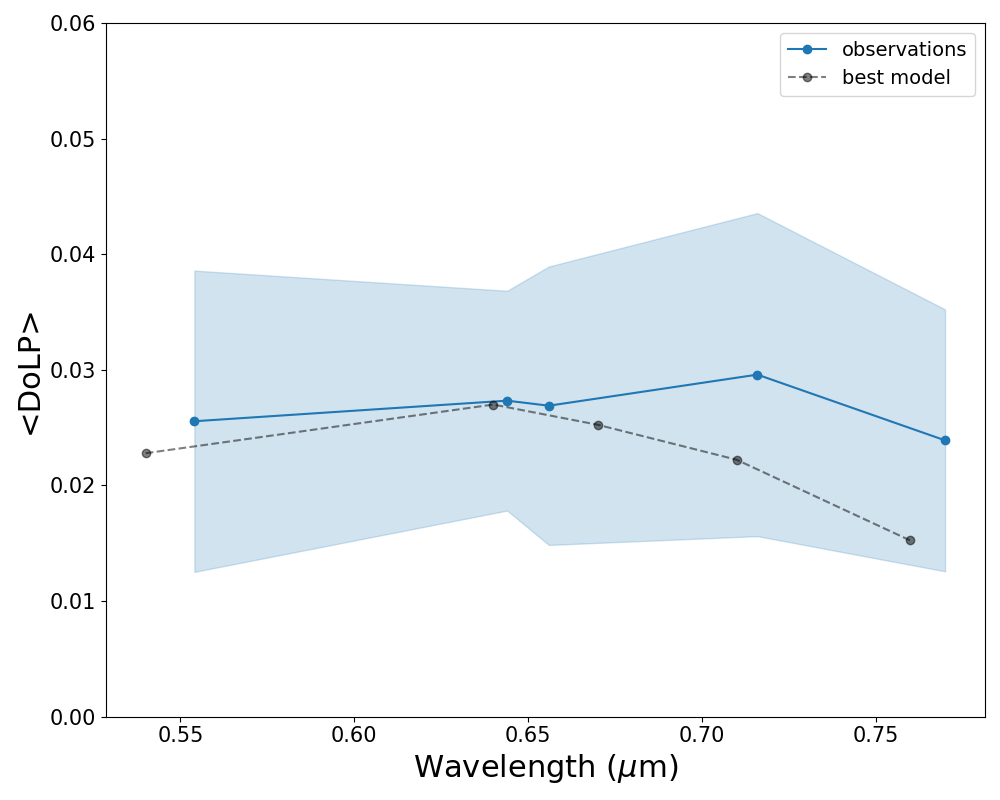}
      \caption{The average DoLP within the southern clump of our best fit model (see Table~\ref{model_params}) as a function of wavelength. Blue circles and blue homogeneous background denote the observations and their uncertainties.}
    \label{dolp_wave}
   \end{figure}

Figure \ref{model} shows the intensity, polarised flux and DoLP for the best fitting model. A side-by-side comparison of the observations and best fitting model can be seen in Fig. \ref{sidebyside}. At longer wavelengths the DoLP diminishes relative to the observations. However, the fit remains within the uncertainties of the (polarised)-flux calibration (see Fig.~\ref{dolp_wave}).

\begin{table}
\caption{Summary of adopted stellar parameters and fitted clump parameters. \label{model_params}}
 \begin{tabular}{lrcl}
    \hline \hline\\[-9pt]
    Parameter             & Value   & Confidence interval  & Unit \\
    \hline\\[-9pt]
    $T_{\rm eff}$       & 3500      &                & K \\
    $g$                 &    1      &                & cm/s$^{2}$ \\
    $R_{\star}$         &  680      &                & R$_{\odot}$  \\
    \hline\\[-9pt]
    $x$                 &  -0.3      &                & $R_{\star}$ \\
    $y$                 &  -4.4      &  -5.0 $-$ -3.2       & $R_{\star}$ \\
    $z$                 &  -2.5     &  -3.1 $-$ -1.3     & $R_{\star}$ \\
    $d_{\rm clump}$     &  5.1     &  4.1 $-$ 5.2       & $R_{\star}$ \\
    $\rho_{\rm clump}$     &  1.07    &  0.70 $-$ 1.10       & $10^{-18}$\,g\,cm$^{-3}$ \\
    $R_{\rm clump}$        &  3.8      &  2.6 $-$ 4.4       & $R_{\star}$ \\
    $M_{\rm clump}$        &  1.3       &  0.3 $-$ 1.5       & $10^{-8}\,$M$_{\odot}$ \\
    \hline
 \end{tabular}
 \vspace{1ex}\\
\begin{flushleft}
\textbf{Note}: Where  $d_{\rm clump}$ is the distance from the centre of the star to the centre of the clump
\end{flushleft}
\end{table}

\section{Discussion}
\label{sec:discussion}
\subsection{Properties of the dusty clump}
\label{sec:cloud_properties}

\begin{figure}   
   \centering
   \includegraphics[width=\hsize]{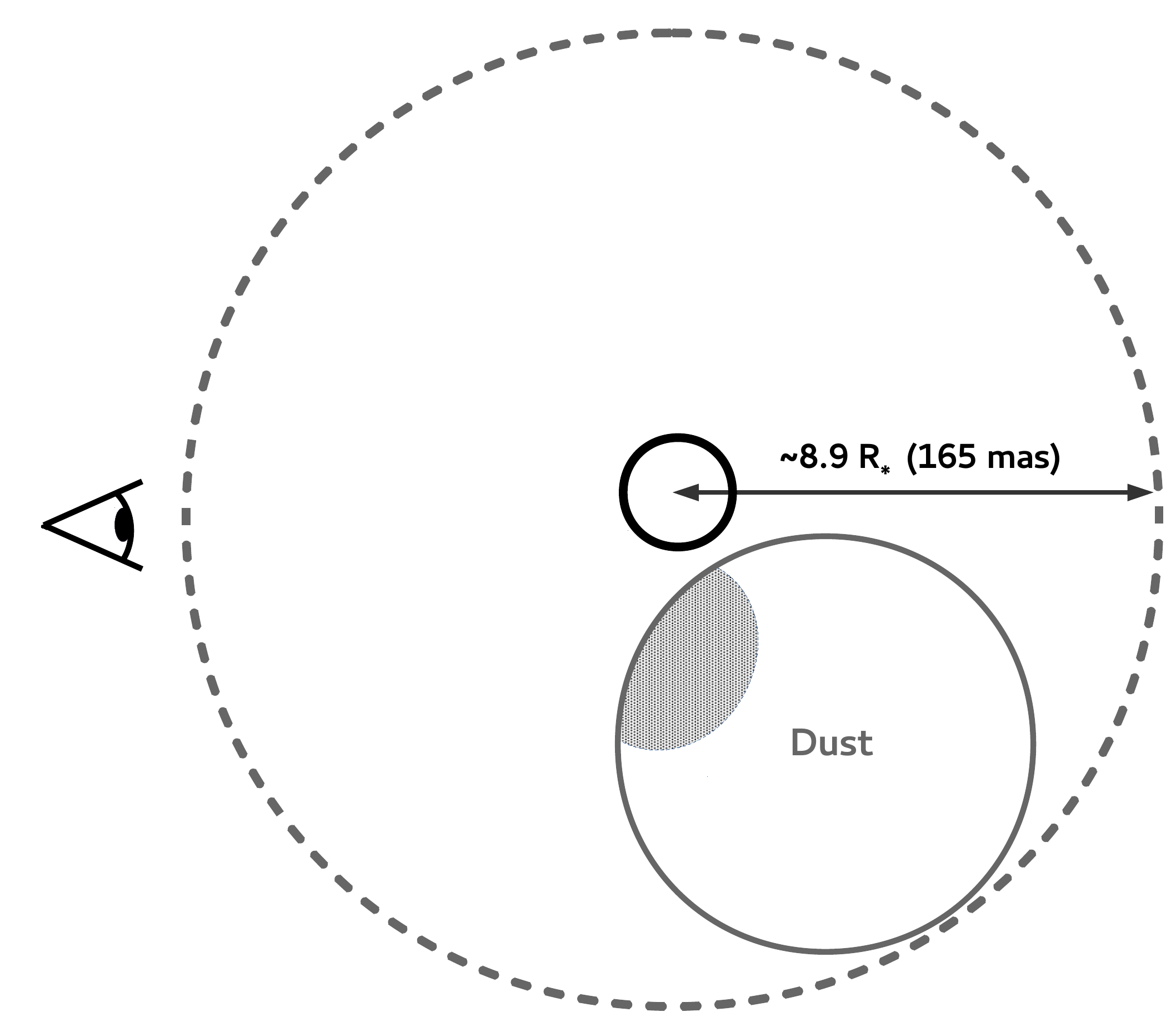}
      \caption{Schematic showing the dust clump and star relative to the direction toward the observer. The furthest edge of the clump from the star is at 165 mas or 8.9\,$R_{\star}$, corresponding to a dynamical flow time from Antares of about 4 years assuming a radial wind speed of 34\,km\,s$^{-1}$. Only part of the clump (approximated by the shaded area) is prominently visible in polarised light. See text for a discussion.}
    \label{dust_schematic}
   \end{figure}

For our homogeneous spherical distribution of dust grains, the best fit to the DoLP images places the centre of the clump at 5.1 $R_{\star}$ and yields a clump radius of 3.8\,$R_{\star}$. Figure ~\ref{dust_schematic} provides a schematic of the geometry of the system. 
The clump is considerably larger than the star itself with the smaller clumps failing to produce as much polarisation signal. The reason for this is the combined effect of the non-isotropic nature of scattering of light off of dust grains (which has a maximum polarisation for 90$^{\circ}$ scattering angle) and the marginally optically thick nature of the clump at optical wavelengths. One therefore most prominently observes polarisation from the part of the clump that is closest to the star. 

The visual extinction in the V-band of a line-of-sight through the centre of the clump is about $A_{\rm V} = 3.6$. A similar clump perfectly aligned in front of the stellar disk would have reduced the visual light flux by a large factor ($\simeq$ 35). This could be an explanation of what has happened to Betelgeuse in 2019-2020, where a significant dimming was observed in visible light (photo release ESO/Montarg\`{e}s et al.\footnote{https://www.eso.org/public/news/eso2003/}; \citealt{2020ApJ...891L..37L}). Given the dimensions of the clump, it fills 18 percent of the total sky as seen from Antares, i.e. a minimum of about 5$-$6 such clump are needed to cover the entire surface.

\citet{2011A&A...526A.156M} assume the gas-to-dust ratio in the circumstellar environment of red supergiants to be $\psi = 200$ in the limit of full condensation of refractory metals. However, a general value of $\psi$ for RSGs still remains highly uncertain. For aluminium oxide the dust temperatures in the clump range from approximately 40 to 1000\,K.
In Sect.~\ref{sec:dust_composition} we pointed out that about 8\% of the clump is, however, too close to the surface to allow silicate or carbon-based grains to survive. Assuming rapid formation (for time-scales refer to \citealt{2019MNRAS.489.4890B}) of these species once the local temperature drops below the condensation temperature as the clump moves away from the surface, we derive a mean gas-to-dust ratio in the clump using $1/\psi = 0.92/200 + 0.08/7000$, i.e. $\psi \simeq 215$, applying 7000 for the gas-to-dust ratio of fully condensed alumina in a solar abundance mixture \citep{2009ARA&A..47..481A} and gas-to-dust ratio of 200 for the rest of the clump. Using a total mass of grains of $1.3 \times 10^{-8}\,$M$_{\odot}$, this implies a total mass of the clump of $\sim 2.8 \times 10^{-6}\,$M$_{\odot}$.

We lack velocity information for the conspicuous nearby dusty clump. Obtaining such kinematic information through resolved spectroscopy is crucially important for identifying the launching and/or driving mechanism of the clump and the time of ejection. Using a speculative velocity of 30\,km\,s$^{-1}$ (inspired from findings by \citealt{2014A&A...568A..17O}; see below), which is lower than the local escape velocity at $r = 5.11\,R_{\star}$ of 43 km\,s$^{-1}$, the clump would have been ejected about 2 years prior to our observations. Note that within this hypothesis the clump would fall down on the star and sublimate, if the radiative pressure on the newly formed dust does not succeed in accelerating it further.

\subsection{Dusty clumps further away from Antares}
\label{sec:furtherclouds}
\citet{2014A&A...568A..17O} use VLT/VISIR at a spatial resolution of 0.5" to probe the $6" \times 6"$ surroundings of Antares in the Q1 filter at 17.7\,$\mu$m. They identify six dusty clumps (at distances ranging from 40 - 96 R$_{\star}$), as well as unresolved emission from the innermost region -- i.e. the region that is probed here in more detail. Assuming the dust consists of an astronomical silicate mixture \citep{2007ApJ...657..810D} they find typical dust masses of their resolved clumps of $(3 - 6) \times 10^{-9}\,$M$_{\sun}$, hence total clump masses of $(0.6 - 1.2) \times 10^{-6}\,$M$_{\sun}$ adopting $\psi = 200$. These values are similar to the total mass we derive for the near-surface clump that is scrutinised here.

The clumps studied by \citet{2014A&A...568A..17O} are spatially unresolved or only marginally resolved, therefore we cannot directly compare clump sizes. Assuming all clumps have a similar size when ejected by the star, the VISIR image implies that internal expansion may have increased their radii by at most a factor of a few. By comparing to an earlier image of Antares' nearby environment by \citet{2001ApJ...548..861M}, taken with the MIRLIN focal-plane array camera at Keck\,II, \citet{2014A&A...568A..17O} estimate the clumps to move out with projected velocities of 13\,$-$\,40 km s$^{-1}$ which the authors conclude is not consistent with a simple monotonically accelerating outflow.

\subsection{Modes and potential driving mechanisms of mass loss from Antares}

\citet{2012A&A...546A...3B} combine measurements of notably Zn\,{\sc ii} absorption line strengths in the line of sight toward the companion Antares B and hydrodynamical simulations of the way in which the B2.5\,V star creates density perturbations in a radial wind from Antares, as well as an H\,{\sc ii} region, to derive the mass loss in gas. They find $\dot{M} = (2.0 \pm 0.5) \times 10^{-6} $\,M$_{\odot}$\,yr$^{-1}$.

How does this total mass loss rate compare to a mean mass loss in ejected dusty clumps? \citet[][see Sect. \ref{sec:furtherclouds}]{2014A&A...568A..17O} estimated that the 6 clumps they observed around Antares to have a mean projected outward directed dusty clump velocity of 34 km\,s$^{-1}$, implying a dynamical crossing time of the projected zone from the closest observed clump to the star to the most distant (40 - 96 R$_{\star}$) of 24.7 years. This yields a typical clump ejection timescale of $\sim$5 yrs and a mean mass loss rate in clumps of approximately $1.5 \times 10^{-7}$\,M$_{\odot}$\,yr$^{-1}$ adopting $\psi = 200$. However, this is a lower limit to the clump mass-loss as it relies on a velocity estimation based on the clump motion in the VISIR field of view. The latter captures only the displacement in the plane of the sky and neglects the motion in the line of sight. Consequently, the actual velocity of each clump could be higher leading to a larger mass-loss rate. Additionally the adopted gas-to-dust ratio of 200 is highly uncertain which would significantly affect our calculation of the total mass lost through clumps. 

The clump in this present study had likely not yet been ejected at the time of \citeauthor{2014A&A...568A..17O}'s 2010 observations, however, its dust mass is comparable to those of the outer clumps studied by \citet{2014A&A...568A..17O}. Taken at face value, the findings of \citeauthor{2014A&A...568A..17O} and the present results suggest that the mass-loss in ejected clumps contributes significantly to the total mass-loss. To provide insight on whether it represents the main mass-loss mechanism or is one of more contributors would require further knowledge of the 3D kinematics of the dusty clumps.

At present, we may only speculate as to the mechanism ejecting clumps of material from the surface of Antares. Variability in both light and radial velocity reveals two preferred characteristic timescales for the star, one of 100\,$-$\,350 days and one of about 6 yrs (e.g. \citealt{2010ApJ...725.1170S}; \citealt{2013AJ....145...38P,2013ApJ...777...10P}). The former has been associated with the typical lifetime of convective cells at the surface \citep[e.g][]{2011A&A...535A..22C} and with fundamental mode or first-order overtone pulsations \citep[e.g.][]{1969ApJ...156..541S}. The latter may be connected to stochastic oscillations, presumably due to the interaction of convection and pulsations \citep{2006MNRAS.372.1721K}, or to the turnover time of convective motions \citep{2010ApJ...725.1170S}. The typical timescale for clump ejection seems to agree best with the latter, longer timescale. 3D hydrodynamical simulations show that the surface of red supergiant stars are covered by a few large convective cells only \citep[e.g.][]{2015ASPC..497...11C} with timescales dependent on the size and depth of the cell \citep{1975ApJ...195..137S}. If gas would be released over the full extent of such a cell -- provided conditions were right -- the surface covering factor would be in line with the 18 per cent derived for the clump that is studied here. Given that not each and every surfacing convective cell releases a cloud of gas, suitable conditions for launching material may depend on the interplay of the multiple processes, likely including convection and pulsations.

\begin{figure*}
      \includegraphics[width = \textwidth]{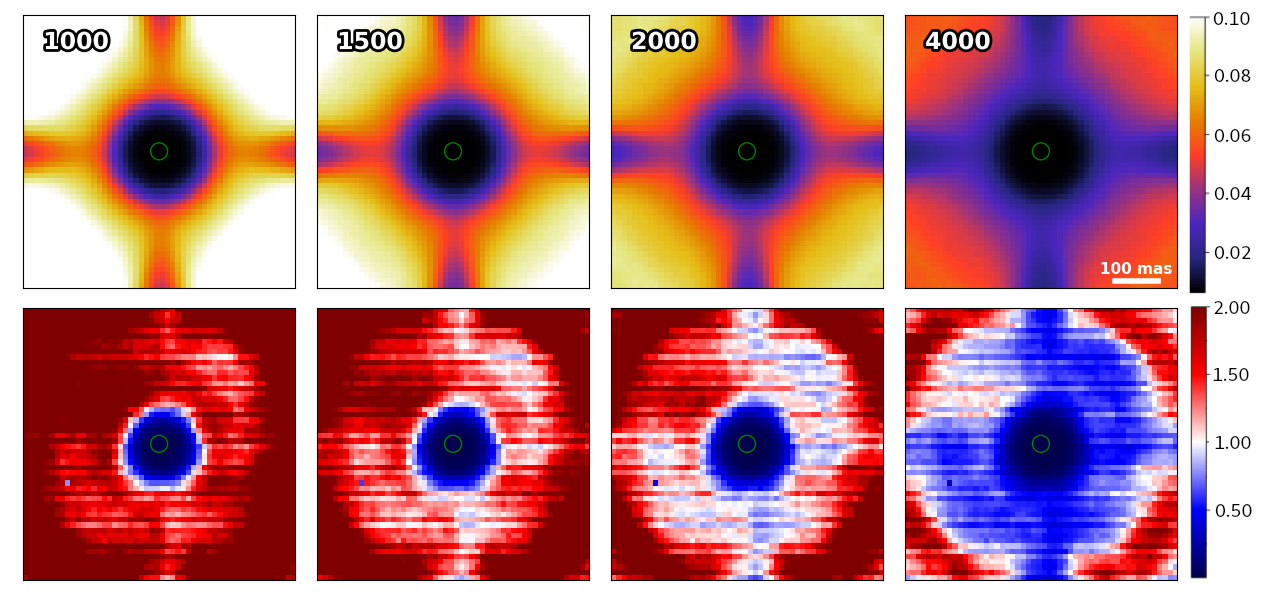}
      \caption{Models of the DoLP produced for a dusty radial outflow in the V-filter. The mass loss rate is set at $2 \times 10^{-6}\,$M$_{\odot}$\,yr$^{-1}$ and the terminal flow velocity at 30\,km\,s$^{-1}$. Dust condensation is assumed to start at 5\,$R_{\star}$. From left to right, the gas-to-dust ratio is 1000, 1500, 2000, and 4000. Top panels show the predicted DoLP in the model. Bottom panels show the predicted DoLP divided by the RMS map of the observations. Blue indicates a less than 1$\sigma$ detection; red a detection that is above 1$\sigma$. As we do not detect significant polarisation in the region outside of the clump, this yields a lower limit to the gas-to-dust ratio of the radial outflow in this region of $\psi_\text{min}$ = 2000.}
         \label{outflow}
   \end{figure*}

Does dust also form in the (initially) radial outflow from Antares as e.g. probed and modelled by \citet{2012A&A...546A...3B}? The top four panels of Fig.~\ref{outflow} show predictions of the DoLP for a radial flow from the star with a mass-loss rate of $2 \times 10^{-6}\,$ \msunyr and a terminal velocity of 30\,km\,s$^{-1}$. The gas-to-dust ratio is set to $\psi = 1000, 1500, 2000$ and 4000 (gas-to-dust ratios below these values show a clear detection in the SPHERE images), and silicate dust is assumed to form instantaneously at 5\,R$_{\star}$. Silicate dust is chosen for the outflow as we expect the lower temperature at this distance to allow for the condensation of less temperature sensitive dust species. The bottom panels show the DoLP predicted by these models divided by the root mean square (RMS) map of the observations. The models were run for the V-filter as it has the lowest uncertainties (being the widest of the six filters used). In these bottom plots the blue areas correspond to where the detection is below 1$\sigma$ and the red areas are above 1$\sigma$. We conclude that the minimum gas-to-dust ratio is $\psi_{\rm min} \sim 2000$ given this mass loss rate, as otherwise we would have detected a polarised signal at the 1$\sigma$ level.

These results clearly point to at most partial dust formation in the radially streaming wind from the star. Dust nucleation computations by \citet{2019MNRAS.489.4890B} indicate a critical density $n_{\rm H} \sim 5 \times 10^{10}$\,g\,cm$^{-3}$ for dust nucleation to set in a galactic environment. This is about three orders of magnitudes larger than the density at 5\,R$_{\star}$ in Antares's flow of $2 \times 10^{-6}\, $ \msunyr, and may help explain the high value for the lower limit $\psi_{\rm min}$ that we report here. \citet{2016A&A...594A.108H} constrain the minimum density of dust grains in AGB outflows for a dust-driven outflow to develop at $n_{\rm d} \geq 4 \times 10^{-6}$\,cm$^{-3}$. For a population of 0.1\,$\mu$m silicate grains this converts to a maximum gas-to-dust ratio for dust driving to occur of $\psi_{\rm max-dd} \sim 500$. Given their higher luminosities, this value may be somewhat higher for RSGs. Still, the derived $\psi_{\rm min}$ and estimated $\psi_{\rm max-dd}$ seem to suggest that the amount of solid state material that may actually form in the radial outflow from Antares is too little to efficiently power a dust-driven wind. If so, the radial outflow requires an altogether different driving mechanism. However, it should be noted that our observations and models here can only account for the inner wind (as pictured in Fig. \ref{outflow}) and may not be representative of what is happening further out from the star. Modelling of ISO-SWS spectra of Antares by \citet{1999A&A...345..605J} also gives a high minimum gas-to-dust ratio in the wind of 600. Dust driving may then only be relevant for the episodic ejection of clumps of gas, in which apparently dust is condensing efficiently, possibly because associated shocks and turbulent eddies produce significant small-scale over-densities.

\subsection{Comparison to other RSGs}

Antares is not the only RSG that has shown evidence for a more inhomogeneous wind. The few RSGs of which the circumstellar environment has been studied show a wide range of characteristics possibly connected to their evolutionary stage. The RSG VY CMa has shown extreme episodic mass-loss (\citealt{2015A&A...584L..10S}; \citealt{2019A&A...627A.114K}). A study by \citet{2019A&A...627A.114K} using ALMA has shown a dusty envelope containing several large clumps. Modelling of the observations suggests that it is these clumps -- being episodically expelled into the interstellar medium -- that are responsible for the high mass-loss (instead of being a result of a steady spherical wind). NOEMA observations of the CO J=2-1 line \citep{2019MNRAS.485.2417M} of the RSG $\mu$ Cep also suggest that the ejection of clumps from the circumstellar environment is a large contributor ($\geq$ 25\%) to the mass-loss. SPHERE/ZIMPOL observations of Betelgeuse \citep{2016A&A...585A..28K} also show a patchy and clumpy inner circumstellar environment and a departure from spherical symmetry in the visible.The clumpy nature of the environment of Betelgeuse has also been observed out to tens of stellar radii with VLT/VISIR by \citet{2011A&A...531A.117K}. A study of the variations of the silicate feature by comparison of IRAS LRS spectra and other ground-based spectra of RSGs spanning 25 years by \citet{1999ApJ...521..261M} shows that the mass-loss characteristics and dust signatures can vary over both short and long timescales. Multiple modes of mass loss, i.e. a clumpy and dusty episodic mass loss and a dust-poor radial outflow of gas, therefore is likely a general phenomenon among RSGs. Which of these modes is dominant may depend on stellar properties and surface conditions

\section{Summary and conclusions}
\label{sec:summary}

The SPHERE/ZIMPOL observations of Antares show a strong localised signal in the DoLP which indicates the presence of a large dusty clump in the inner circumstellar environment. Modelling the observations using the radiative transfer code \textsc{MCMax3D} shows that the clump is 2.6 - 4.4 $\text{R}_{\star}$ in size with a dust mass of about $(0.3 - 1.5) \times 10^{-8}$ \msun. Our models place the edge of the clump beyond the plane of the sky through the centre of Antares (so, the dusty clump is `behind' the star) and its inner edge within 0.5 $ \text{R}_{\star}$ from the stellar surface. Adopting full condensation of solids in the dusty clumps ($\psi \sim 200$) and incorporating findings by \citet{2014A&A...568A..17O}, we find a minimum mass-loss rate from clumps of $1.5 \times 10^{-7}$  \msunyr. No significant polarisation is measured in the rest of the probed ambient environment, placing constraints on the abundance of dust in a radially streaming stellar wind. Using the canonical value for this radial mass loss of $2 \times 10^{-6}\,$ \msunyr \citep{2012A&A...546A...3B} the gas-to-dust ratio in this flow must be at least $\psi_{\rm min} = 2000$ in the field of view of ZIMPOL. This suggests that the inner region of the radial flow is not dust driven.

The surface covering factor of the dusty clump (18\%) agrees quite well with the expected size of surfacing convective cells in RSGs. Moreover, the estimated typical ejection timescale for clumps (of $\sim$5 yrs; \citealt{2014A&A...568A..17O}) matches well with a characteristic timescale for photometric and radial velocity variability (of $\sim$6 yrs) that has been associated to (an interplay of) pulsational and convective behaviours. This points towards convection and pulsation playing a role in the launching mechanism of the dusty clumps. The methodology developed here, i.e. to constrain the 3D position of recently ejected dusty-clumps, in principle allows to empirically study a possible connection of this mode of mass loss with surface activity. This requires long-term simultaneous interferometric monitoring of surface structures and the direct stellar surroundings. Supplementing this with kinematic information through resolved spectroscopy may further aid in establishing or rejecting such a connection.

\section*{Acknowledgements}
The authors acknowledge funding from the KU Leuven C1 grant MAESTRO C16/17/007.
This project has received funding from the European Union's Horizon 2020 research and innovation program under the Marie Sk\l{}odowska-Curie Grant agreement No. 665501 with the research Foundation Flanders (FWO) ([PEGASUS]$^2$ Marie Curie fellowship 12U2717N awarded to M.M.).
L.D. acknowledges support from the ERC consolidator grant 646758 AEROSOL.
This work has made use of the the SPHERE Data Centre, jointly operated by OSUG/IPAG (Grenoble), PYTHEAS/LAM/CESAM (Marseille), OCA/Lagrange (Nice), Observatoire de Paris/LESIA (Paris), and Observatoire de Lyon.
We acknowledge with thanks the variable star observations from the AAVSO International Database contributed by observers worldwide and used in this research.
This research made use of IPython \citep{PER-GRA:2007}, Numpy \citep{5725236}, Matplotlib \citep{Hunter:2007}, SciPy \citep{2020SciPy-NMeth}, Astropy\footnote{Available at \url{http://www.astropy.org/}}, a community-developed core Python package for Astronomy \citep{2013A&A...558A..33A}, and  Uncertainties\footnote{Available at \url{http://pythonhosted.org/uncertainties/}}: a Python package for calculations with uncertainties.

\section*{Data availability}
The data underlying this article are available from the ESO archive under programme ID 095.D-0458.

\bibliographystyle{mnras}
\bibliography{references}




\appendix

\section{}

   \begin{figure*}   
   \centering
   \includegraphics[width=\textwidth]{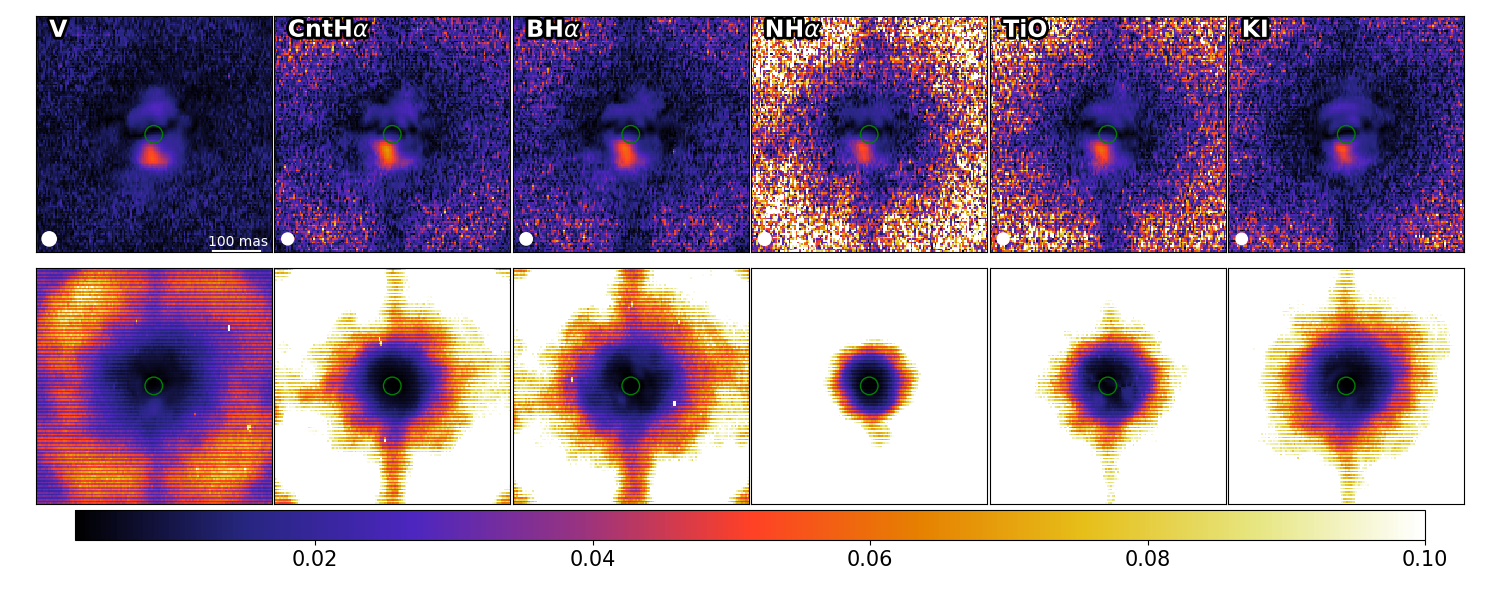}
      \caption{Degree of linear polarisation (top) with corresponding rms map (bottom). Outside of the region of interest discussed in this study the observations become dominated by noise.}
    \label{rms}
   \end{figure*}


\bsp	
\label{lastpage}
\end{document}